\documentclass[pra,twocolumn,superscriptaddress]{revtex4}
\usepackage{amsfonts}
\usepackage{amsmath}
\usepackage{amsthm}
\usepackage{amscd}
\usepackage{amssymb}
\usepackage{subfigure}
\usepackage{amsxtra}
\usepackage{bm}           
\usepackage{bbm}
\usepackage{graphicx}
\usepackage{epstopdf}
\usepackage{color}

\def\bi#1\ei {\begin{itemize}#1\end{itemize}}
\def\bn#1\en {\begin{enumerate}#1\end{enumerate}}
\def\bea#1\eea {\begin{align}#1\end{align}}
\def\bean#1\eean {\begin{align*}#1\end{align*}}
\def\ben#1\een {\begin{equation*}#1\end{equation*}}
\def\be#1\ee {\begin{equation}#1\end{equation}}
\def\bes#1\ees {\begin{equation}\begin{split}#1\end{split}\end{equation}}
\def\bear#1\eear {\begin{eqnarray}#1\end{eqnarray}}
\def\bear#1\eear {\begin{eqnarray*}#1\end{eqnarray*}}

\def\red#1{\textcolor{red}{#1}}

\newcommand{\beq}{\begin{equation}}
\newcommand{\eeq}{\end{equation}}

\newcommand{\ketbra}[2]{\ensuremath{| #1 \rangle \langle #2 |}}

\begin{document}
\title{Secret key expansion from covert communication}
\author{Juan Miguel Arrazola}
\affiliation{Centre for Quantum Technologies, National University of Singapore, 3 Science Drive 2, Singapore 117543}
\author{Ryan Amiri}
\affiliation{SUPA, Institute of Photonics and Quantum Sciences, Heriot-Watt University, Edinburgh EH14 4AS, United Kingdom}
\begin{abstract}
Covert communication allows us to transmit messages in such a way that it is not possible to detect that the communication is occurring. This provides protection in situations where knowledge that people are talking to each other may be incriminating to them. In this work, we study how covert communication can be used for a different purpose: secret key expansion. First, we show that any message transmitted in a secure covert protocol is also secret and therefore unknown to an adversary. We then propose a protocol that uses covert communication where the amount of key consumed in the protocol is smaller than the transmitted key, thus leading to secure secret key expansion. We derive precise conditions showing that secret key expansion from covert communication is possible when there are sufficiently low levels of noise for a given security level. We conclude by examining how secret key expansion from covert communication can be performed in a computational security model. 
\end{abstract}
\maketitle

Quantum cryptography encompasses a wide range of protocols that employ principles of quantum mechanics to secure communication.  Quantum key distribution (QKD) is a method for key expansion where two parties with access to an insecure quantum channel and a small shared secret key can securely generate a much larger shared secret key. This new secret key can then be used be used to encrypt communications or to generate more secret key \cite{scarani2009security,diamanti2016practical}. Besides QKD, there are many other protocols in quantum cryptography that are beginning to shift from theoretical proposals to experimental demonstrations. Examples of these are quantum signature schemes \cite{AWKA2016,AWA2015,amiri2015unconditionally,collins2017experimental,collins2016experimental,yin2017experimental}, quantum fingerprinting \cite{arrazolaqfp,xu2015experimental,GX16}, bit commitment \cite{ng2012experimental,lunghi2013exp,liu2014experimental,verbanis201624}, and quantum money \cite{W1983,PYJ2011,Gav2012,GK2015,amiri2017quantum,bartkiewicz2017experimental,bozzio2017experimental}. 

Covert communication offers a method to transmit messages while hiding the fact that the communication is happening at all. This allows users to prevent adversaries from detecting their transmissions, providing a method to secure their messages even in cases where revealing that they are talking to each other may be incriminating to them. Recently, a square-root law for covert communication has been proven, stating that $O(\sqrt{N})$ covert bits can be reliably sent over $N$ channel uses, even in the presence of an unbounded quantum adversary\cite{bash2013limits,bash2013quantum,che2013reliable,bash2015hiding,wang2016fundamental,sheikholeslami2016covert,
arumugam2016keyless,bloch2016covert,bash2016covert,wang2016optimal,bash2015quantum}. Quantum communication protocols can also be carried out covertly, either through noisy channels as in the classical case \cite{arrazola2016covert} or by exploiting relativistic quantum effects \cite{bradler2016absolutely,bradler2017covert}. 

In this work, we study how the techniques of covert communication can be employed to achieve secure secret key expansion. We show that the security statement of covert communication is formally equivalent to composable security for the secrecy of the transmitted messages. This implies that all information sent in a secure covert communication protocol is unknown to an eavesdropper and can therefore be used as a secret key. Covert communication protocols require an initial shared secret key between the participants, which raises the question of whether more secret key can be obtained than is consumed. We describe a covert communication protocol that is capable of generating more secret key than is consumed, thus constituting a method for secure and covert secret key expansion. We conclude by discussing how secrecy can be achieved and quantified in a computational model where pseudorandomness is employed, which is likely to be necessary for practical realizations of covert communication.

\section{Secrecy from covert communication} Alice wants to transmit a message to Bob in such a way that an eavesdropper Eve cannot distinguish whether they are communicating or not. To quantify Eve's ability to detect the communication, we assume that Alice either communicates or not with equal probability and Eve's goal is to correctly distinguish between these two scenarios. Eve's detection error probability $P_e$ is given by $P_e=\frac{1}{2}(P_{FA}+P_{MD})$, where $P_{FA}$ is the probability of a false alarm and $P_{MD}$ is the probability of a missed detection. The goal of a secure protocol is to prevent Eve from performing better than a random guess, i.e. it must be that $P_e\geq \frac{1}{2}-\epsilon$ for sufficiently small $\epsilon>0$. We refer to $\epsilon$ as the detection bias. 

Covert communication requires noise in the channel connecting Alice and Bob, who have access to $N$ optical modes that can be used by Alice to transmit signals to Bob. Eve has access to these modes once they leave Alice's lab and can perform any operation on them to attempt to detect the communication. In all known covert communication protocols, Alice and Bob achieve covertness by using a shared secret key to randomly spread their signals among the $N$ available modes, allowing them to ``hide" their transmissions in the channel's noise.  

We can formally define the security of a covert communication protocol in terms of a state discrimination problem. When Alice and Bob are not talking to each other, Eve receives a state $\rho$ which depends on the properties of the noise. When they do communicate, Eve gets a state $\sigma$ that depends on both the noise and the signals sent by Alice. The protocol is secure if Eve cannot distinguish these two states. From Helmstrom's bound \cite{helstrom76a}, we can relate the detection bias $\epsilon$ to the trace distance between $\rho$ and $\sigma$ as
\beq
\frac{1}{2}||\rho-\sigma||_1=2\epsilon.
\eeq
A covert communication protocol is $\epsilon$-secure if this relation holds for any message sent by Alice to Bob.
 
In quantum key distribution (QKD), security of a secret key generated from a QKD protocol is also quantified in terms of the trace distance between two states: the shared state $\rho_{KE}$ between Alice, Bob and Eve after carrying out the protocol, and the ideal state $\red{\rho_U}\otimes \rho_E$, where the quantum state $\red{\rho_U}$ describes a uniformly distributed key that is completely decoupled from Eve's state $\rho_E$. More precisely, the key generated by a QKD protocol is called $\varepsilon$-secure if
\beq
\frac{1}{2}||\rho_U\otimes \rho_E-\rho_{KE}||_1\leq \varepsilon.
\eeq 
This definition is composable, meaning that such a secret key can be used as an input to other protocols without compromising their security. 

To connect this definition to the security of covert communication, we include reference to an additional quantum system held by Alice corresponding to her choice of message. In that case, from Eve's perspective, the state when no covert communication takes place is given by
\beq
\rho'=\sum_{m}p(m)\ketbra{m}{m}\otimes \rho,
\eeq
where $m$ denotes the possible messages, $p(m)$ is the probability that it is selected for transmission and $\rho$ is the noisy state as before. Here, for convenience, we imagine that Alice has selected a message $m$ with probability $p(m)$ but simply does not transmit it. The state $\rho'$ is thus an ideal state for secrecy of the message since Eve is completely uncorrelated from message register, and would be so for any choice of state in Alice's register other than $\sum_{m}p(m)\ketbra{m}{m}$, so we do not lose generality by making this choice. Similarly, when Alice communicates with Bob, the total state is given by
\beq
\sigma'=\sum_{m}p(m)\ketbra{m}{m}\otimes\sigma_{m},
\eeq
where $\sigma_m$ is the state of the modes when message $m$ is sent. If the covert communication protocol is $\epsilon$-secure, by definition it holds that
\beq
\frac{1}{2}||\rho-\sigma_m||=2\epsilon
\eeq
for all $m$. We then have that
\begin{align*}
||\rho'-\sigma'||_1&=||\sum_{m}p(m)\ketbra{m}{m}\otimes(\rho-\sigma_{m})||_1\\
&\leq \sum_{m}p(m)||\ketbra{m}{m}\otimes (\rho-\otimes\sigma_{m})||_1\\
&=\sum_{m}p(m)||\rho-\sigma_{m}||_1\\
&\leq \sum_{m}p(m)4\epsilon=4\epsilon,
\end{align*}
where we have used convexity of the trace norm. Therefore, for any $\epsilon$-secure covert communication protocol it holds that
\beq
\frac{1}{2}||\rho'-\sigma'||\leq 2\epsilon.
\eeq
We can conclude that if a covert communication protocol is $\epsilon$-secure, then the transmitted message is $2\epsilon$-secure with respect to its secrecy. This holds even when the message is covertly transmitted as plaintext, i.e. without any form of encryption, meaning that a secret key can be distributed using a covert communication protocol by sending a uniformly random string as the message. However, as discussed before, all known covert communication protocols require a shared secret key between Alice and Bob. To achieve key expansion, a covert communication protocol must therefore transmit more secret key than it consumes. In the following, we give an explicit example of such a key expansion protocol.

\section{Key expansion protocol} 

\begin{figure}[t!]
\includegraphics[width=8cm]{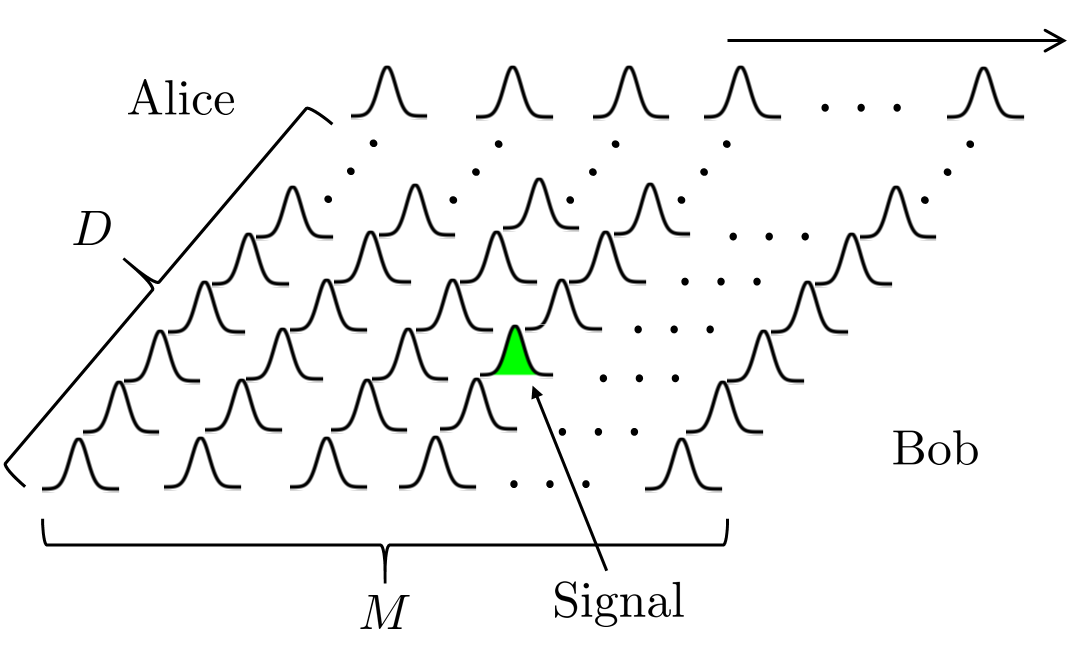}
\caption{(Color online) Covert communication protocol for secret key expansion. Alice and Bob have $N=MD$ modes at their disposal, which they divide into $M$ blocks of $D$ modes each. They randomly select one of the blocks and they send one signal chosen uniformly at random among all the $D$ modes, allowing them to retrieve $\log D$ bits of key in an ideal case. To do so, they require $\log M$ bits of pre-shared secret key, so the protocol generates more key than it consumes as long as $D>M$. In practice, we must also account for the cost of error-correction, leading to more sophisticated conditions for key expnasion.}\label{Fig:Data1}
\end{figure}

Alice and Bob have $N=MD$ modes at their disposal, which they divide into $M$ blocks of $D$ modes each. Their goal is to covertly transmit a uniformly random string that is longer than the secret key they require to carry out the protocol. To do so, they first use their secret key to select one of the $M$ blocks uniformly at random, which requires a key of $\log M$ bits. Here and throughout the paper, all logarithms are in base 2. Once the block is selected, they transmit a random string of $\log D$ bits by selecting one of the $D$ modes uniformly at random and sending a single-photon signal in it. From Eve's perspective, this strategy is equivalent to sending a signal uniformly at random in one of the $N$ available modes.  Let $\rho_n$ be Eve's state for a single mode when there is only noise present and let $\rho_s$ be the state of a single mode when a single photon is sent. When there is no communication, Eve's state is tensor product state given by
\beq
\rho=\rho_n^{\otimes N}.
\eeq
On the other hand, when Alice and Bob do communicate, Eve's state is given by
\beq
\sigma=\frac{1}{N}\sum_{i=1}^N\Pi_i(\rho_s\otimes\rho_n^{\otimes N-1}),
\eeq
where $\Pi_i$ is a permutation that places the signal state $\rho_s$ in the $i$-th mode. The detection bias can be calculated in terms of the trace distance between these two states, which is challenging to calculate given that $\sigma$ is not a tensor product state. Instead, we want to relate the detection bias $\epsilon$ of this protocol to the detection bias $\epsilon_{I}$ of an i.i.d. protocol where Alice chooses whether to send a signal in each mode with equal probability. These protocols have been carefully studied in the literature \cite{bash2015quantum,arrazola2016covert} and closed forms exist for their detection biases.

In the i.i.d. protocol, let $p_k$ be the probability that Alice sends $k$ signals. Conditioned on her sending $k$ signals, there is a resulting detection bias $\epsilon_k$ for Eve which can be used to express the actual detection bias $\epsilon_{I}$ as
\beq\label{epsPrime}
\epsilon_{I}=\sum_{k=0}^N p_k \epsilon_k.
\eeq
Now, by definition $\epsilon_0=0$, and $\epsilon_1=\epsilon$ since exactly one signal chosen uniformly at random among all modes is sent in the original protocol. Assume, for the purpose of contradiction, that $\epsilon_1>\frac{\epsilon_{I}}{1-p_0}$. Since $\epsilon_{k+1}>\epsilon_k$ for all $k$, it holds that
\begin{align*}
\sum_{k=0}^N p_k \epsilon_k&=\sum_{k=1}^N p_k \epsilon_k\\
&>\sum_{k=1}^N p_k\left(\frac{\epsilon_{I}}{1-p_0}\right)=\epsilon_{I}, \\
\end{align*}
which is in contradiction with \eqref{epsPrime}. Therefore, we conclude that
\beq
\epsilon\leq \frac{\epsilon_{I}}{1-p_0}.
\eeq
This expression allows us to relate the detection bias of the key expansion protocol to known detection biases of protocols previously studied in the literature. Note that in general both $\epsilon_{I}$ and $p_0$ depend on the protocol being considered. For i.i.d. protocols, a square-root law has been proven for the detection bias as a function of the total number of time-bins \cite{bash2015quantum,arrazola2016covert} stating that
\beq
\epsilon_{I}\leq \frac{\beta}{\sqrt{N}}
\eeq
for some constant $\beta$ that depends on the protocol and its parameters. This can be used to bound the detection bias of the original protocol as
\beq\label{Eq:DetBias}
\epsilon\leq \frac{\beta}{(1-p_0)\sqrt{N}}.
\eeq
The remaining challenge is to determine the conditions for which more secret key can be transmitted than what is consumed.

For a given $\epsilon$, we can use Eq. \eqref{Eq:DetBias} to calculate the smallest $N=MD$ that achieves the desired bound. This fixes the number of blocks to be
\beq\label{Eq:MandD}
M=\left(\frac{\beta}{\epsilon(1-p_0)}\right)^2 \frac{1}{D}.
\eeq
We require $\log M$ secret bits to specify which block was chosen. For the protocol to transmit more secret bits than it consumes we need to have $D>M$, which leads to the condition
\beq\label{Eq: ineq1}
D>\frac{\beta}{\epsilon(1-p_0)}.
\eeq
From this it is clear that we want to make $D$ as large as possible. However, there is a limit which arises from the reliability of decoding the message. In the presence of noise, there is a probability $p_c$ of observing at least one photon in any mode due to the noise present. If Bob observes a click in a mode different than the one where the signal was sent, he will obtain an error. The probability $\delta$ of obtaining at least one click in one of the other $D-1$ modes is
\beq\label{Eq: delta}
\delta(D)=1-(1-p_{c})^{D-1}.
\eeq
We can model the effect of error due to noise in the channel as a symmetric channel for $D$ symbols that either transmits the correct symbol with probability $1-\delta$ or changes it to another one of the $D-1$ remaining ones with probability $\delta$. The capacity of this channel, which quantifies the number of bits that can be reliably transmitted per use of the channel, is given by \cite{weidmann2012fresh}
\beq\label{Eq:Capacity}
C=\log D-h[\delta(D)]-\delta(D)\log(D-1),
\eeq
where $h(p)=-p\log(p)-(1-p)\log(1-p)$ is the binary Shannon entropy. From this it is clear that we require $\delta(D)$ to be small in order to achieve reliability of the transmission. From Eq. \eqref{Eq: delta}, this implies the condition $Dp_c\ll 1$, which sets a limit to the size of $D$. Overall, more secret key will be produced than consumed as long as $C>\log M$. Combining Eqs. \eqref{Eq: ineq1} and \eqref{Eq:Capacity} while using the approximations $\log(D-1)\approx \log D$ we arrive at the key expansion condition
\beq
[2-\delta(D)]\log D>2\log\left(\frac{\beta}{\epsilon(1-p_0)}\right)+h[\delta(D)].
\eeq
Whether there exists a value of $D$ for which this condition can be met depends ultimately on the noise level -- which fixes the parameters $\beta$ and $p_c$ -- and on the target detection bias $\epsilon$. The i.i.d. protocol that is used to bound the detection bias can be optimized to fix the value of $p_0$ and $\beta$ for a given noise level. In the following, we discuss an explicit family of protocols that can meet this condition for key expansion.

We focus on protocols that use a single-photon signal to transmit information and where noise originates from background thermal radiation with mean photon number per mode $\bar{n}$. As shown in Ref. \cite{arrazola2016covert}, for $\bar{n}\ll 1$, the detection bias of an i.i.d. protocol can be bounded as
\beq
\epsilon_{I}\leq\frac{d}{4}\frac{1}{\sqrt{2N \bar{n}}},
\eeq
where $d=qN$ is the average number of signals sent in the i.i.d. protocol. Note that we have that 
\beq
\beta=\frac{d}{4\sqrt{2\bar{n}}}
\eeq
in this case. The probability of not sending any signals in the i.i.d. protocol is
\begin{align*}
p_0&=(1-q)^N\\
&=\left(1-\frac{d}{N}\right)^N\approx e^{-d},
\end{align*}
where the last approximation holds in the limit of $N\gg 1$. Finally, the probability $p_c$ of observing at least one photon in each of the noisy modes is given by
\beq
p_c=1-\frac{1}{1+\bar{n}}\approx\bar{n}
\eeq
which leads to a value of 
\beq\label{Eq: delta2}
\delta(D)=1-\left(\frac{1}{1+\bar{n}}\right)^{D-1},
\eeq
for the error probability. Plugging in these expressions we obtain the key expansion condition
\beq
[2-\delta(D)]\log D>2\log\left(\frac{d}{4\sqrt{2\bar{n}}(1-e^{-d})\epsilon}\right)+h[\delta(D)].
\eeq
For given values of $\bar{n}$ and $\epsilon$, it suffices to fix a value of $d$ for the i.i.d. protocol and determine whether there exist values of $D$ such that this inequality is satisfied. However, we can significantly simplify the condition by using a series of approximations. 

First, we fix the value of $d$ such that $d/(4\sqrt{2}(1-e^{-d}))=1$. Similarly, we consider only values of $D$ such that $D\bar{n}\ll 1$ and we ignore any term that is linear in $D\bar{n}$. This leads to a simplified key expansion condition
\beq\label{Eq:SimpleCondition}
2\log D>\log\left(\frac{1}{\bar{n}\epsilon^2}\right).
\eeq
From this, together with the condition $D\bar{n}\ll 1$, it is easy to see that the key expansion condition can be achieved as long as $1/\epsilon^2$ is small compared to $1/\bar{n}$. Moreover, from Eq. \eqref{Eq:MandD}, the total number of modes needed is approximately $N=1/(\bar{n}\epsilon^2)$. 

\begin{figure}[t!]
\includegraphics[width=\columnwidth]{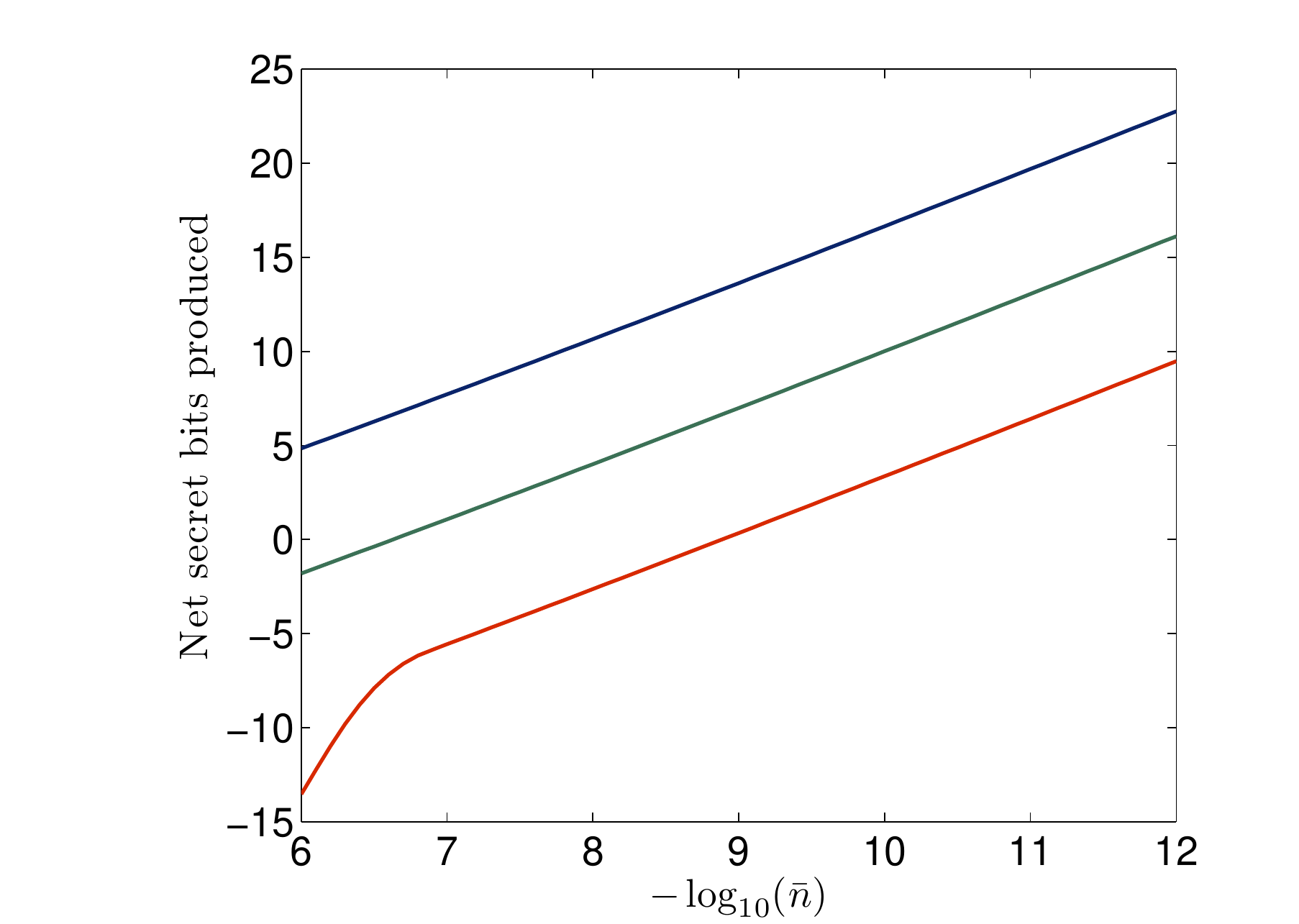}
\caption{(Color online) Number of secret bits covertly produced per run of the protocol, i.e. the difference between secret bits transmitted and secret bits consumed, as a function of the noise parameter $\bar{n}$. The top curve (blue) corresponds to a security level of $\epsilon=0.1$, the middle curve (green) to $\epsilon
=0.01$ and the bottom curve (red) to $\epsilon=0.001$. A negative value implies that more secret bits are consumed than generated.}\label{Fig:BitsProduced}
\end{figure}

For given values of $\bar{n}$ and $\epsilon$, we can simply compute the value of $D$ that leads to the largest difference between the number of secret bits transmitted and the secret bits consumed. This is illustrated in Fig. \ref{Fig:BitsProduced}, where we plot the net number of secret bits, i.e. the difference between secret bits transmitted and secret bits consumed, that can be produced per run of the protocol for different values of the noise level $\bar{n}$ and the security level $\epsilon$. From Fig. \ref{Fig:BitsProduced}, we learn that the ideal regime to conduct secret key expansion corresponds to low noise levels and that a significant price is paid for increasing security. Indeed, under the assumptions used to derive Eq. \eqref{Eq:SimpleCondition}, if we fix $D=\alpha/\bar{n}$ for some $\alpha < 1$, we can approximate the net number of bits produced $K$ as
\beq
K\approx(1-\alpha)\log\left(\frac{1}{\bar{n}}\right)+(2-\alpha)\log(\alpha)-2\log\left(\frac{1}{\epsilon}\right),
\eeq
showing how that the net number of secret bits produced increases with small $\bar{n}$ but decreases with small $\epsilon$, in accordance with the exact calculations used for Fig. \ref{Fig:BitsProduced}.

In conclusion, we have formally shown that covertness also provides secrecy of the transmitted message, which allows secret keys to be transmitted securely using covert communication. We have also shown that in principle it is possible to build covert communication protocols that transmit more secret key than they consume, leading to a novel method of key expansion that is also covert. In a practical setting, transmission losses will make it very challenging to transmit more secret bits than are consumed, although it could potentially be achieved over short distances using brighter coherent-state signals instead of single photons and efficient detectors. 

An alternative method for using covert communication for secret key expansion is to consider a computational security model where pseudorandom number generators (PRNGs) are used to decide in what blocks to send the covert signals. This has the advantage that a small amount of initial secret key can be used to covertly transmit a much larger amount of new secret key. In the next section, we discuss how composable security can also be obtained in this computational model and provide explicit protocols for secret key expansion.

\section{Key expansion in a computational model}
Pseudorandom number generators (PRNGs) are algorithms that take a truly random bit string -- referred to as the seed -- and output a much larger bit string. The goal of a PRNG is to create an output that cannot be distinguished from a truly random string by a computationally bounded adversary. Formally, let $l$ be a polynomial and let $f:\{0,1\}^n\rightarrow \{0,1\}^{l(n)}$ be a PRNG mapping seeds of $n$ bits to pseudorandom strings of $l(n)$ bits, with $l(n)>n$. The pseudorandom function $f$ is $\delta(n)$-secure if for any polynomial-time algorithm $D$ it holds that \cite{katz2014introduction}
\beq
|\Pr[D(f(s))=1]-\Pr[D(r)=1]|\leq \delta(n),
\eeq
where $r$ is a uniform random string of $l(n)$ bits. The probabilities are taken over a uniform choice of $s$, $r$ and any randomness in the algorithm $D$. It is usually required that $\delta(n)$ is an exponentially small function of $n$. Note that this definition is also composable as it is stated in terms of distinguishing the output of a PRNG from the ideal case of a truly random string. 

In covert communication, instead of using a truly random secret string to choose where to send signals, Alice and Bob can instead use the key as the seed of a PRNG, employing its pseudorandom output to decide where to send the signals. This is in fact what is done in practice in frequency-hopping and spread spectrum communications \cite{simon1994spread}. In the following, we describe a secure key expansion protocol in a computational model.

\subsection{Protocol}

\begin{enumerate}
\item Alice and Bob share an $n$-bit secret key $k_0$ as well as the pseudorandom output $f(k_0)$ produced from a publicly known PRNG $f$ with security $\delta(n)$.
\item Alice uniformly at random selects a new $m$-bit key $k_1$.
\item To transmit $k_1$, she first uses an error-correcting code $E$ to produce a codeword $E(k_1)$ of the key. She uses the pseudorandom output $f(k_0)$ to choose the modes used to covertly transmit $E(k_1)$ to Bob. 
\item Following transmission, Bob holds $E(k_1)'$, which may not be identical to $E(k_1)$.
\item Bob performs error correction to recover $k_1$. 
\end{enumerate}

If the covert communication protocol is $\epsilon$-secure, this implies, as discussed previously, that the transmitted secret key is $2\epsilon$-secure with respect to secrecy from an adversary -- but only when true randomness is employed. If a PRNG is used instead, then by definition a polynomial adversary cannot distinguish the PRNG output from true randomness except with probability $\delta(n)$, which is exponentially small in $n$. From the union bound, this implies that the detection bias of the covert protocol is at most $\epsilon+\delta(n)$, leading to a $2(\epsilon+\delta(n))$-security of the new secret key. Note that these statements are independent of the particular covert communication protocol employed as long as it is $\epsilon$-secure.

This key expansion protocol can produce more secret key than it consumes as long as $m>n$, i.e. as long as the size of the transmitted message is larger than the PRNG seed. Most of the widely used PRNGs such as the Advanced Encryption Standard (AES) and the Secure Hashing Algorithm (SHA-3) have outputs that are exponentially larger than their seeds while still retaining computational indistinguishability from true randomness, which itself is exponentially small in the input seed. This shows that a very large key can be expanded using covert communication in a computational security model.

In this sense, covert communication can serve a similar role as PRNG encryption, where security is obtained by performing a modulo 2 sum of the message with a PRNG output, as is commonly done with AES. In both cases, a small secret key can be employed to protect a large amount of communication, but in the case of covert communication the transmission is also protected from detection by an adversary. These two methods could even be combined for additional security by covertly transmitting an encrypted key.

\section{Conclusion}
We have shown that it is in principle possible to perform secret key expansion using covert communication, even in the presence of unbounded quantum adversaries. We have given conditions for when key expansion is possible, and shown that they can be met whenever the noise level is sufficiently small for a given security level. From a fundamental point of view, this is a novel method of performing secure secret key expansion where noise is not a hindrance but a vital resource. In practice, meeting the required conditions for key expansion in the presence of imperfections will be challenging, especially considering the inherent low rates of covert communication. Hence it is likely to be preferable to employ pseudorandomness to determine in which modes the covert signals are sent. Our results indicate that there is a vital benefit to covert communication besides its direct goal of hiding communication from adversaries -- it can also serve as a powerful method for transmitting data requiring the highest levels of security.

\bibliography{Bibliography}
\bibliographystyle{apsrev}
\end{document}